Hindawi

*Review Article*

# A Survey of Quantum Lyapunov Control Methods

**Shuang Cong and Fangfang Meng**

*Department of Automation, University of Science and Technology of China, Hefei 230027, China*

Correspondence should be addressed to Shuang Cong; scong@ustc.edu.cn

The condition of a quantum Lyapunov-based control which can be well used in a closed quantum system is that the method can make the system convergent but not just stable. In the convergence study of the quantum Lyapunov control, two situations are classified: nondegenerate cases and degenerate cases. For these two situations, respectively, in this paper the target state is divided into four categories: the eigenstate, the mixed state which commutes with the internal Hamiltonian, the superposition state, and the mixed state which does not commute with the internal Hamiltonian. For these four categories, the quantum Lyapunov control methods for the closed quantum systems are summarized and analyzed. Particularly, the convergence of the control system to the different target states is reviewed, and how to make the convergence conditions be satisfied is summarized and analyzed.

## 1. Introduction

The theory of quantum mechanics is one of the major discoveries in the history of science in the 20th century. It is a very important issue to study the properties of the quantum mechanical systems and their control. According to whether the system is isolated or not, a quantum mechanical system can be a closed system or an open system. In a closed quantum system, the evolution of the state is unitary. There are mainly two methods to describe the evolution of a closed quantum system's states. They are the Schrödinger equation $i|\dot{\psi}(t)\rangle = (H_0 + \sum_{k=1}^{r} H_k u_k(t))|\psi(t)\rangle$ and the quantum Liouville equation $i\dot{\rho}(t) = [H_0 + \sum_{k=1}^{r} H_k u_k(t), \rho(t)]$, in which $|\psi(t)\rangle$ is the quantum state vector, $\rho(t)$ is the density operator, $H_0$ is the internal Hamiltonian, and $H_k, (k = 1, \ldots, r)$ are the control Hamiltonians. In an open quantum system, the system interacts with the surroundings, thus the loss of the system's information leads to the non-unitary evolution of the state. The most common method to describe the open system is the Lindblad master equation: $i\dot{\rho} = [H_0 + \sum_{k=1}^{r} H_k u_k(t), \rho] + L_D(\rho)$, which is in fact the sum of a closed system and a dissipative term caused by the loss of the information or energy. Obviously, the research of the properties of the closed quantum system and its control is relatively simple. Moreover, there is a more important fact: the research of the closed quantum system is the basis of that of the open quantum system.

Quantum control has attracted much attention in recent years and it has been found the potential applications in many fields such as atomic physics [1–4], molecular chemistry [5–9] and quantum information [10, 12]. Up to now, there have been many quantum methods, such as quantum optimal control [13–15], adiabatic control [16–18], the Lyapunov-based control [19–41], and optimal Lyapunov-based quantum control [42]. For the Lyapunov-based quantum control, it is relatively easy to design an analytical but not numerical control law, and the control system based on this control method is at least stable, so it has been a common control method.

One of the major concerns of the Lyapunov control is choosing an appropriate Lyapunov function to design the control laws. Ordinarily, the control laws and the control effects are different when the Lyapunov functions are distinct. It's a good idea to choose the Lyapunov function based on the geometrical and physical meanings. Usually, there are mainly three Lyapunov functions to be selected: the Lyapunov function based on the state distance [19–21, 26–32, 39–41], the state error [22, 23, 37, 38], and the average value of an imaginary mechanical quantity [24, 25, 39, 41]. The so-called imaginary mechanical quantity means that it is a linear Hermitian operator to be designed and may be not a physically meaningful observable quantity such as coordinate and energy. Among these three Lyapunov functions, the Lyapunov-based quantum control methods based on the state distance and state error only need to adjust the scale factors of the control laws. These two Lyapunov control methods are relatively simple and easy to grasp. The Lyapunov-based quantum control based on the average value of an imaginary mechanical quantity contains more adjustable parameters. So it is more flexible and also more complex at the same time.

Generally speaking, the Lyapunov-based control method can only ensure that the control system is stable. The probability control in the quantum system requires us to design a control strategy which can make the system convergent, because a stable quantum control method may result in that the control system cannot reach the desired target state. Therefore another major concern of this control strategy for the closed quantum systems is the convergence of the control systems. So far, there have been the following research results on the convergence of the closed quantum systems [19, 23–25, 27–29, 33, 36].

(I) For the Schrödinger equation, the convergence conditions are as follows. (i) The internal Hamiltonian is strongly regular; (ii) All the eigenstates, which are different from the target state, are directly coupled to the target state for the Lyapunov control based on the state distance or the state error, or any two eigenstates are coupled directly for the Lyapunov control based on the average value of an imaginary mechanical quantity.

(II) For the quantum Liouville equation, the convergence conditions are as follows. (i) The internal Hamiltonian is strongly regular, and (ii) the control Hamiltonians are full connected.

At first, the Lyapunov control method which could only ensure the convergence to the target eigenstate was studied [19, 23, 25]. Then, the target mixed state, which commutes with the internal Hamiltonian, was studied [33, 36]. Later, the convergence to the target superposition state was solved by means of designing the control laws in the interaction picture of the control system, and the convergence to the target mixed state which does not commute with the internal Hamiltonian was solved by using a unitary transformation.

In fact, many actual systems do not satisfy the convergence conditions mentioned above, such as the time domain model of the selective excitation of the stimulated Raman scattering [43], the coupled two spin systems, and one-dimensional oscillators [44]. These systems are called non-ideal systems and in the degenerate cases. For the degenerate cases, the convergence of the control systems was solved by introducing a series of implicit function perturbations and choosing an implicit Lyapunov function [37–40]. At first, the convergence to the target eigenstate was only guaranteed [37–40]. Then the convergence to the target mixed state which commutes with the internal Hamiltonian was solved by introducing a series of implicit function perturbations, and the convergence to the target superposition state and the target mixed state which does not commute with the internal Hamiltonian was solved by introducing a series of constant disturbances.

The aim of this paper is to summarize and analyze the existing Lyapunov control methods for the nondegenerate and degenerate cases, respectively. Dividing the target state into four categories: the eigenstate, the mixed state which commutes with the internal Hamiltonian, the superposition state, and the mixed state which does not commute with the internal Hamiltonian state, we summarize the design methods of the control laws, analyze the convergence to the target state, and investigate how to make these conditions of the convergence be satisfied.

The remainder of this paper is arranged as follows. In Section 2, the research results for the nondegenerate cases are summarized and analyzed. In Section 3, the research results for the degenerate cases are summarized and analyzed. Some concluding remarks are drawn in Section 4.

## 2. Non-Degenerate Cases

The design of the control laws and the analysis of the convergence are very important in the Lyapunov control method. The design of the control laws is based on the Lyapunov stability theorem, which is to design the control laws to make the selected positive semi-definite Lyapunov function $V(t)$ satisfy $\dot{V}(t) \leq 0$. The convergence analysis of this control method is mainly based on the LaSalle invariance principle [44] for the autonomous systems, or the improved Barbalat lemma [45] for the non-autonomous systems.

In this Section, we will summarize and analyze the research results on the convergence of the control system in the non-degenerate cases for the target state being an eigenstate, a mixed state which commutes with the internal Hamiltonian, the superposition state, or/and a mixed state which does not commute with the internal Hamiltonian state, respectively.

*2.1. Target Eigenstate.* It is convenient to use the bilinear Schrödinger equation to describe the control systems if the target state is a pure state. Consider the $N$-level closed quantum system governed by the following bilinear Schrödinger equation:

$$ i \left| \dot{\psi}(t) \right\rangle = \left( H_0 + \sum_{k=1}^{r} H_k u_k(t) \right) \left| \psi(t) \right\rangle, \tag{1} $$

where $|\psi(t)\rangle$ is the quantum state vector, $H_0$ is the internal Hamiltonian, $H_k, (k = 1, \ldots, r)$ are control Hamiltonians, and $u_k(t), (k = 1, \ldots, r)$ are scalar and real control laws.

In the following sections, in the case that the target state $|\psi_f\rangle$ is an eigenstate, that is, $H_0|\psi_f\rangle = \lambda_f|\psi_f\rangle$, where $\lambda_f$ is the eigenvalue of the internal Hamiltonian $H_0$, the research results on the convergence are summarized and analyzed for the Lyapunov control based on the state distance, the state error, and the average value of an imaginary mechanical quantity, respectively.

*2.1.1. Lyapunov Control Based on State Distance.* Consider the following Lyapunov function based on the state distance:

$$ V_1 \left( |\psi\rangle \right) = \frac{1}{2} \left( 1 - \left| \left\langle \psi \mid \psi_f \right\rangle \right|^2 \right). \tag{2} $$

The time derivative of the Lyapunov function (2) is

$$\dot{V}_1 = -\sum_{k=1}^{r} u_k(t) \cdot \text{Im} \left[ \left\langle \psi \mid \psi_f \right\rangle \left\langle \psi_f \mid H_k \mid \psi \right\rangle \right]. \quad (3)$$

The control laws $u_k(t)$ which can make $\dot{V}_1 \le 0$ hold can be designed as

$$u_k(t) = K_k f_k \left( \text{Im} \left( \left\langle \psi \mid \psi_f \right\rangle \left\langle \psi_f \mid H_k \mid \psi \right\rangle \right) \right), \quad (k = 1, \ldots, r), \quad (4)$$

where $K_k > 0$, and $y_k = f_k(x_k)$, $(k = 1, \ldots, r)$ are monotonic increasing functions through the coordinate origin of the plane $x_k - y_k$.

The control laws designed in (4) can only ensure the control system (1) to be stable. One needs to do further study on the convergence of the control system. The control system governed by (1) is an autonomous system, whose convergence can be analyzed based on the LaSalle invariance principle [44]. According to the LaSalle invariance principle, as $t \rightarrow \infty$, any state trajectory will converge to the largest invariant set contained in the set $E$ in which the states satisfy that the first order derivative of the Lyapunov function equals zero. In fact, the set $E$ contains not only the target state but also other states, thus the system may converge to other states rather than the target state. The main idea to solve this problem is to add restrictions to make the set $E$ as small as possible. Based on the LaSalle invariance principle, the convergence of the control system governed by (1) can be depicted by Theorem 1.

**Theorem 1** (see [19]). *Consider the control system governed by* (1) *with control fields* $u_k(t)$ *designed in* (4). *If (i) The system is strongly regular, that is,* $\omega_{i'j'} \neq \omega_{lm}$, $(i', j') \neq (l, m)$, $i', j', l, m \in \{1, 2, \ldots, N\}$, $\omega_{lm} = \lambda_l - \lambda_m$, *where* $\lambda_l$ *is the lth eigenvalue of* $H_0$ *corresponding to the eigenstate* $|\phi_l\rangle$; *(ii) for any* $|\phi_i\rangle \neq |\psi_f\rangle$, $i \in \{1, \ldots, N\}$, *there exists at least a* $k \in \{1, \ldots, r\}$ *such that* $\langle \phi_i | H_k | \psi_f \rangle \neq 0$, *then any state trajectory will converge toward* $E_1 = \{e^{i\theta} | \psi_f \rangle, \theta \in R\}$.

From Theorem 1, one can see that in the case that the target state is an eigenstate, if the control system governed by (1) satisfies the conditions (i)-(ii), the control system can converge to the equivalent state of the target eigenstate $e^{i\theta} |\psi_f\rangle$ from any initial pure state. These two conditions are relevant to the internal Hamiltonian and the control Hamiltonians, which are system parameters. Once the control system is determined, the Hamiltonians are fixed and cannot be changed by designing the control laws.

*2.1.2. Lyapunov Control Based on State Error.* Consider the following Lyapunov function based on the state error:

$$V_2 \left( |\psi\rangle \right) = \frac{1}{2} \left\langle \psi - \psi_f \mid \psi - \psi_f \right\rangle. \quad (5)$$

In the case of selecting the Lyapunov function based on the state error defined by (5), in order to facilitate to design the control laws based on the Lyapunov stability theorem, the drift item appeared in the first order time derivative of Lyapunov function, which is caused by the internal Hamiltonian, is needed to be eliminated. The existing solution is to add a global phase control item $\omega$ into the control system governed by (1). This method will not change the population distribution of the control system. Thus the dynamical equation (1) becomes

$$i \left| \dot{\psi}(t) \right\rangle = \left( H_0 + \sum_{k=1}^{r} H_k u_k(t) + \omega I \right) |\psi(t)\rangle. \quad (6)$$

After some deduction, one can obtain the time derivative of the Lyapunov function (5) as

$$\dot{V}_2 = -\left( \lambda_f + \omega \right) \Im \left( \left\langle \psi_f \mid \psi \right\rangle \right) - \sum_{k=1}^{r} \Im \left( \left\langle \psi_f \mid H_k \mid \psi \right\rangle \right) u_k(t). \quad (7)$$

The control laws which can make $\dot{V}_2 \le 0$ hold can be designed as

$$\omega = -\lambda_f + c f_0 \left( \Im \left( \left\langle \psi_f \mid \psi \right\rangle \right) \right), \quad (8)$$

$$u_k(t) = K_k f_k \left( \Im \left( \left\langle \psi_f \mid H_k \mid \psi \right\rangle \right) \right), \quad (k = 1, \ldots, r), \quad (9)$$

where $K_k > 0$, and $y_k = f_k(x_k)$, $(k = 0, \ldots, r)$ are the monotonic increasing functions through the coordinate origin of the plane $x_k - y_k$.

Based on the LaSalle invariance principle, the convergence of the control system governed by (6) can be depicted by Theorem 2.

**Theorem 2** (see [19, 23]). *Consider the control system governed by* (6) *with control fields* $u_k(t)$ *designed in* (9) *and* $\omega$ *designed in* (8). *If (i)* $\omega_{i'j'} \neq \omega_{lm}$, $(i', j') \neq (l, m)$, $i', j', l, m \in \{1, 2, \ldots, N\}$, $\omega_{lm} = \lambda_l - \lambda_m$, *where* $\lambda_l$ *is the lth eigenvalue of* $H_0$ *corresponding to the eigenstate* $|\phi_l\rangle$; *(ii) for any* $|\phi_i\rangle \neq |\psi_f\rangle$, $i \in \{1, \ldots, N\}$, *there exists at least a* $k \in \{1, \ldots, r\}$ *such that* $\langle \phi_i | H_k | \psi_f \rangle \neq 0$. *Then any state trajectory will converge toward* $E_2 = \{e^{i\theta} | \psi_f \rangle, \theta \in R\}$.

From Theorem 2, one can see that for the case that the target state is an eigenstate, if the control system governed by (6) satisfies the conditions (i)-(ii), the control system can also converge to the equivalent state of the target eigenstate $e^{i\theta} |\psi_f\rangle$ from any initial pure state.

*2.1.3. Lyapunov Control Based on Average Value of an Imaginary Mechanical Quantity.* Consider the following Lyapunov function based on the average value of an imaginary mechanical quantity:

$$V_3 \left( |\psi\rangle \right) = \left\langle \psi | P | \psi \right\rangle, \quad (10)$$

where the imaginary mechanical quantity $P$ is a positive definite Hermitian operator.

The first order time derivative of the Lyapunov function (10) can be obtained as

$$\dot{V}_3 = i \left\langle \psi \right| \left[ H_0, P \right] |\psi\rangle + i \sum_{k=1}^{r} \left\langle \psi \right| \left[ H_k, P \right] |\psi\rangle u_k. \quad (11)$$

Set $[H_0, P] = 0$ such that the drift term in the right side of (11) can be eliminated. In order to ensure $\dot{V}_3 \leq 0$, one can design $u_k(t)$ as

$$u_k(t) = -K_k f_k \left( i \left\langle \psi \right| [H_k, P] \left| \psi \right\rangle \right), \quad (k = 1, \ldots, r), \quad (12)$$

where $K_k > 0$, and $y_k = f_k(x_k), (k = 1, \ldots, r)$ are monotonic increasing functions through the coordinate origin of the plane $x_k - y_k$.

Then based on the LaSalle invariance principle, all the state trajectories of the system will converge to the invariant set contained in the set $E$ in which $\dot{V} = 0$ holds. Denote the state at the time $t$ as $|\psi(t)\rangle = \sum_{l=1}^N c_l(t)|\phi_l\rangle$, where $c_l(t)$ is the coefficient corresponding to the $l$th eigenstate $|\phi_l\rangle$. For the Schrödinger equation governed by (1), if the control system is strongly regular and any eigenstate is directly coupled to all other eigenstates, that is, for any $j \neq l$, $j, l \in \{1, 2, \ldots, N\}$, there exists a $k \in \{1, \ldots, r\}$ such that $\langle \phi_j | H_k | \phi_l \rangle \neq 0$, then one can deduce that $\dot{V}_3(t) = 0$ holds for all $t \geq t_0, t_0 \in R$ is equivalent to

$$\left( P_l - P_j \right) c_j(t_0) c_l^*(t_0) = 0, \quad (l, j = 1, \ldots, N), \quad (13)$$

where $P_l$ and $P_j$ are the $l$th and $j$th eigenvalues of $P$, respectively.

If the target state $|\psi_f\rangle$ is an eigenstate, and all the eigenvalues of $P$ are designed mutually different, that is, $P_l \neq P_j$ for any $l \neq j$, $(l, j = 1, 2 \ldots, N)$, then (13) is equivalent to

$$c_j c_l^* = 0, \quad (l, j = 1, \ldots, N). \quad (14)$$

Equation (14) implies that there is at most one $c_j \neq 0, (j = 1, \ldots, N)$; that is, the system will converge to an eigenstate with $t \to \infty$. Thus the convergence of the control system (1) can be depicted by Theorem 3.

**Theorem 3** (see [19, 24, 25]). *Consider the control system governed by (1) with the control fields $u_k(t)$ designed in (12). If (i) $\omega_{i'j'} \neq \omega_{lm}, (i', j') \neq (l, m), i', j', l, m \in \{1, \ldots, N\}$, $\omega_{lm} = \lambda_l - \lambda_m$, where $\lambda_l$ is the $l$th eigenvalue of $H_0$ corresponding to the eigenstate $|\phi_l\rangle$; (ii) for any $i \neq j$, $i, j \in \{1, \ldots, N\}$, there exsits at least a $k \in \{1, \ldots, r\}$ such that $\langle \phi_i | H_k | \phi_j \rangle \neq 0$; (iii) $[H_0, P] = 0$; (iv) for any $l \neq m \in \{1, 2, \ldots, N\}$, $P_l \neq P_m$ holds, where $P_l$ is the $l$th eigenvalue of $P$. Then any state trajectory will converge toward $E_3 = \{e^{i\theta}|\phi_i\rangle\}, i \in \{1, \ldots, N\}, \theta \in R\}$.*

From Theorem 3, one can see that the control system will converge from any initial pure state to an eigenstate which may not be the target eigenstate. In order to make the system converge to the target eigenstate $|\psi_f\rangle$ from any initial pure state $|\psi_0\rangle$, as $\dot{V}_3(t) \leq 0$, one can add a restriction as

$$V_3 \left( \left| \psi_f \right\rangle \right) < V_3 \left( \left| \psi_0 \right\rangle \right) < V_3 \left( \left| \psi_{\text{other}} \right\rangle \right), \quad (15)$$

where $|\psi_0\rangle$ is the initial state, $|\psi_{\text{other}}\rangle$ represents any other state in the set $E_3$ except the target state.

In such a way, any state trajectory of the system will converge to $e^{i\theta}|\psi_f\rangle$ from any initial pure state $|\psi_0\rangle$.

Next, let us analyze how to make these convergence conditions be satisfied. Conditions (i) and (ii) are only relevant to the internal Hamiltonian $H_0$ and the control Hamiltonians $H_k, (k = 1, \ldots, r)$ which cannot be changed by designing appropriate control laws. Condition (iii) means that $P$ and $H_0$ have the same eigenstates. In order to make condition (iii) be satisfied, the eigenvalues of $P$ can be designed as

$$P = \sum_{j=1}^N P_j \left| \phi_j \right\rangle \left\langle \phi_j \right|. \quad (16)$$

Because $P$ should be positive definite, one needs to design $P_i > 0, (i = 1, \ldots, N)$. The restriction (15) can be satisfied by means of designing an appropriate $P$. For the restriction (15), Grivopoulos and Bamieh proposed a design principle of $P$ to make $V_3(|\psi_f\rangle) < V_3(|\psi_{\text{other}}\rangle)$ hold. This design principle of $P$ can be depicted by Proposition 4.

**Proposition 4** (see [25]). *With the constraint condition $\langle \psi | \psi \rangle = 1$, the set of critical points of the Lyapunov function $V_3(|\psi\rangle) = \langle \psi | P | \psi \rangle$ is given by the normalized eigenvectors of $P$. The eigenvectors with the largest eigenvalue are the maxima of $V_3$, the eigenvectors with the smallest eigenvalue are the minima and all others are saddle points.*

According to Proposition 4, in order to make $V_3(|\psi_f\rangle) < V_3(|\psi_{\text{other}}\rangle)$ hold, $P_i > P_f, (i = 1, \ldots, N, P_i \neq P_f)$ needs to be designed, where $P_f$ is the eigenvalue of $P$ corresponding to $|\psi_f\rangle$. Then let us consider the whole restriction (15); it is an attraction problem. If the eigenvalues of $P$ except $P_f$ are close together, the attraction region will be very large. For the limiting case $P_i = P_j > P_f > 0, (i \neq j, P_i, P_j \neq P_f)$, the attraction region will be the whole state space. Thus the design principle of $P$ is $P_i > P_f, (i = 1, \ldots, N, P_i \neq P_f)$ and to make $P_i, P_j \neq P_f$ close together.

In conclusion, for the target state being an eigenstate, the design principle of $P$ is as follows:

(i) $P_l \neq P_m > 0$ for any $l \neq m \in \{1, 2, \ldots, N\}$,

(ii) $P_i > P_f, (i = 1, \ldots, N, P_i \neq P_f), P_i, P_j \neq P_f$ are close together,

(iii) Equation (16): $P = \sum_{j=1}^N P_j |\phi_j\rangle\langle\phi_j|$.

From the above analyses, we can conclude that the Lyapunov control method based on the imaginary mechanical quantity proposed in [19, 24, 25] can only ensure the convergence to an eigenstate, but cannot guarantee the convergence to the target eigenstate from any initial pure state. However, if there exists a $P$ to make the restriction (15) hold, then any state trajectory of the system will converge to the equivalent state of the target eigenstate $e^{i\theta}|\psi_f\rangle$ from any initial pure state.

*2.1.4. Relations between Three Lyapunov Functions.* In the Liouville space, the Hilbert-Schmidt distance between two density operators $\rho_1$ and $\rho_2$ is

$$d_{HS}(\rho_1, \rho_2) = \sqrt{\mathrm{tr}(\rho_1 - \rho_2)^2}. \tag{17}$$

The inner product of two operators $A$ and $B$ is defined as $\langle\langle A|B\rangle\rangle = \mathrm{tr}(A^\dagger B)$, where the operation $A^\dagger$ refers to the conjugate transpose of $A$. Because $\rho = |\psi\rangle\langle\psi|$, the square of the Hilbert-Schmidt distance between the density operator $\rho$ and the target density operator $\rho_f$ can be deduced as

$$d_{HS}^2(\rho, \rho_f) = 2\left(1 - \left|\langle\psi \mid \psi_f\rangle\right|^2\right). \tag{18}$$

One can conclude from (18) that the Lyapunov function based on the state distance $V_1(|\psi\rangle) = (1/2)(1 - |\langle\psi \mid \psi_f\rangle|^2)$ and the state error $V_2(|\psi\rangle) = (1/2)\langle\psi - \psi_f \mid \psi - \psi_f\rangle$ are equivalent.

For the Lyapunov function based on the average value of an imaginary mechanical quantity $V_3(|\psi\rangle) = \langle\psi|P|\psi\rangle$, if the imaginary mechanical quantity $P = (1/2)(I - |\psi_f\rangle\langle\psi_f|)$, $V_3(|\psi\rangle)$ need be $V_1(|\psi\rangle)$ [19]. Therefore $V_1(|\psi\rangle)$ is formally a special case of $V_3(|\psi\rangle)$.

In fact, $V_1(|\psi\rangle)$, $V_2(|\psi\rangle)$, and $V_3(|\psi\rangle)$ can be unified in the following quadratic Lyapunov function:

$$V_4(|\psi\rangle) = \left\langle \psi - \alpha\psi_f \mid Q \mid \psi - \alpha\psi_f \right\rangle \tag{19}$$

Next, we consider three cases as follows:

(i) $\alpha = 0, Q = (1/2)(I - |\psi_f\rangle\langle\psi_f|)$, $V_4(|\psi\rangle)$ will reduce to $V_1(|\psi\rangle)$;

(ii) $\alpha = 1, Q = (1/2)I$, $V_4(|\psi\rangle)$ will reduce to $V_2(|\psi\rangle)$;

(iii) $\alpha = 0$, $V_4(|\psi\rangle)$ will reduce to $V_3(|\psi\rangle)$.

We can see that $V_1(|\psi\rangle)$, $V_2(|\psi\rangle)$, and $V_3(|\psi\rangle)$ are special cases of $V_4(|\psi\rangle)$.

From the above analyses, we can conclude that all these three Lyapunov control methods can converge to the equivalent state of the target eigen state from any initial pure state. The Lyapunov control methods based on the state distance and the state error have only one adjustable parameter. So these two methods are very easy to grasp and very simple. The Lyapunov control based on the average value of an imaginary mechanical quantity has more adjustable parameters. So it is more flexible and also more complex at the same time. For the target eigenstate, $V_1(|\psi\rangle)$ and $V_2(|\psi\rangle)$ are equivalent, so the Lyapunov control based on the state distance and the state error have similar control effects. Because $V_1(|\psi\rangle)$ is formally a special case of $V_3(|\psi\rangle)$, generally, the control effect of the Lyapunov control based on the average value of an imaginary mechanical quantity is better than that of the state distance. At least, it can get the same control effect as the Lyapunov control of the state distance.

## 2.2. Target Mixed State Which Commutes with the Internal Hamiltonian.

The bilinear Schrödinger equation cannot describe the mixed state. Thus for the target mixed state,

it needs to use the quantum Liouville equation which can describe the evolution of any state of a closed quantum system. Consider the $N$-level closed quantum system governed by the following quantum Liouville equation:

$$i\dot\rho(t) = \left[ H_0 + \sum_{k=1}^{r} H_k u_k(t), \rho(t) \right], \tag{20}$$

where $\rho(t)$ is the density operator.

Consider the Lyapunov function based on the average value of an imaginary mechanical quantity:

$$V_5(\rho) = \mathrm{tr}(P\rho). \tag{21}$$

By means of setting $[H_0, P] = 0$, the first order time derivative of the Lyapunov function (21) can be deduced as

$$\dot V_5 = -i\,\mathrm{tr}\left([P, H_0]\rho\right) - i\sum_{k=1}^{r}\mathrm{tr}\left([P, H_k]\rho\right)u_k. \tag{22}$$

In order to ensure $\dot V_5 \leq 0$, one can design $u_k(t)$ as

$$u_k(t) = i\varepsilon_k\,\mathrm{tr}\left([P, H_k]\rho\right), \quad (k = 1, \ldots, r), \tag{23}$$

where $\varepsilon_k \in R, \varepsilon_k > 0$.

Next, let us analyze the convergence to the target state. For the control system governed by (20) in the non-degenerate case, one can deduce that $\dot V_5(t) = 0$ holds for all $t \geq t_0, t_0 \in R$ is equivalent to

$$(H_k)_{jl}\left(P_l - P_j\right)\rho_{lj}(t) = 0, \quad j, l = 1, \ldots, N, \ j < l, \tag{24}$$

where $\rho_{lj}(t)$ is the $(l; j)$th element of the state $\rho(t)$. If the target state $\rho_f$ commutes with $H_0$, that is, $[H_0, \rho_f] = 0$, and all the eigenvalues of $P$ are designed mutually different, that is, $P_l \neq P_j$ for any $l \neq j$, $l, j = 1, 2, \ldots, N$, then (24) is equivalent to

$$\rho_{lj} = 0, \quad (l, j = 1, \ldots, N) \tag{25}$$

which implies that the system will converge to a state which commutes with $H_0$. Thus based on the LaSalle invariance principle, the convergence of the control system governed by (20) can be depicted by Theorem 5.

**Theorem 5** (see [37]). *Consider the control system governed by (20) with the control field $u_k(t)$ designed in (23). If (i) the internal Hamiltonian is strongly regular, that is, $\omega_{i'j'} \neq \omega_{lm}, (i', j') \neq (l, m), i', j', l, m \in \{1, \ldots, N\}, \omega_{lm} = \lambda_l - \lambda_m$, where $\lambda_l$ is the $l$th eigenvalue of $H_0$ corresponding to the eigenstate $|\phi_l\rangle$; (ii) the control Hamiltonians are full coupled; that is, $\forall j \neq l$, for $k = 1, \ldots, r$, there exists at least a $(H_k)_{jl} \neq 0$; (iii) $[H_0, P] = 0$; (iv) $P_l \neq P_m$, $l \neq m \in \{1, 2, \ldots, N\}$, where $P_l$ is the $l$th eigenvalue of $P$. Then any state trajectory will converge toward $E_4 = \{\rho \mid (\rho)_{ij} = 0, \ i, j \in \{1, \ldots, N\}\}$, where $(\rho)_{ij}$ is the $(l, l)$th element of $\rho$.*

From Theorem 5, one can see that if the control system satisfies the conditions (i)–(iv), the control system will converge from any initial state to a state that commutes with the internal Hamiltonian, which may not be the target state. Next, what we need to do is to make the control system converge to the target state.

Denote the state in $E_4$ as $\rho_{E_4}$; then $[\rho_{E_4}, H_0] = 0$ holds which implies that $\rho_{E_4}$ and $H_0$ have the same eigenstates. Since the evolution of $\rho(t)$ is unitary, $\rho(t)$ for $t \geq 0$ are isospectral. So the eigenvalues of $\rho_{E_4}$ are a permutation of the eigenvalues of $\rho_0$. Thus the set $E_4$ has countable elements. If the initial state is generic, that is, the eigenvalues of the initial state are mutually different, the set $E_4$ will have $N!$ elements. For the target state $\rho_f$ which commutes with the internal Hamiltonian, that is, $[\rho_f, H_0] = 0$, in order to make the system converge to the target state $\rho_f$ which commutes with $H_0$ from any initial state $\rho_0$, Kuang and Cong proposed a restriction as

$$V_5 \left( \rho_f \right) < V_5 \left( \rho_0 \right) < V_5 \left( \rho_{\text{other}} \right), \qquad (26)$$

where $\rho_{\text{other}}$ represents any other state in the set $E_4$ except the target state $\rho_f$.

In such a way, if there exists a $P$ to make the restriction (26) hold, any state trajectory of the system will converge to the target state $\rho_f$ which commutes with $H_0$ from any initial state $\rho_0$.

For how to make conditions (iii)-(iv) to be satisfied, please read Section 2.1.3. One can deduce the design principle of $P$ such that $V_5(\rho_f) < V_5(\rho_{\text{other}})$, the result can be depicted by Proposition 6.

**Proposition 6.** *If $(\rho_f)_{ii} < (\rho_f)_{jj}$, $1 \leq i, j \leq N$ holds, then one can design $P_i > P_j$; if $(\rho_f)_{ii} = (\rho_f)_{jj}$, $1 \leq i, j \leq N$ holds, then one can design $P_i \neq P_j$; else if $(\rho_f)_{ii} > (\rho_f)_{jj}$, $1 \leq i, j \leq N$ holds, then one can design $P_i < P_j$, thus $V_5(\rho_f) < V_5(\rho_{\text{other}})$ holds, where $(\rho_f)_{ii}$ is the $(i, i)$th element of $\rho_f$.*

It is difficult to design $P$ such that (26) holds for any initial state $\rho_0$ and any target state $\rho_f$ which satisfies $[\rho_f, H_0] = 0$. One possible method is to introduce a series of implicit function perturbations into the control laws, this method will be presented in Section 3.3.

For the target state $\rho_f$ which commutes with the internal Hamiltonian, the design principle of $P$ is as follows:

(i) $P_l \neq P_m > 0$ for any $l \neq m \in \{1, 2, \ldots, N\}$;

(ii) Design $P$ according to Proposition 6;

(iii) Equation (16): $P = \sum_{j=1}^{N} P_j |\phi_j\rangle\langle\phi_j|$.

In conclusion, if the control system satisfies the conditions (i)–(iv) in Theorem 5 and there exists a $P$ to make the restriction (26) hold, any state trajectory of the system will converge to the target mixed state $\rho_f$ which commutes with the internal Hamiltonian from any initial state $\rho_0$.

*2.3. Target Superposition State.* From Section 2.1.3, one can see that for the control system governed by (1), $\dot{V}(t) = 0$ holds for all $t \geq t_0, t_0 \in R$ is equivalent to $(P_l - P_j)c_j c_l^* = 0, (l, j = 1, \ldots, N)$. If the target state is a superposition state as $|\psi_f\rangle = \sum_{d=k1}^{ko} c_d |\phi_d\rangle, (1 \leq k1 < ko \leq N)$, one can design

$P_{k1} = P_{k2} = \cdots = P_{ko}$ and other eigenvalues of $P$ are mutually different, then $(P_l - P_j)c_j c_l^* = 0, (l, j = 1, \ldots, N)$ is equivalent to

$$c_j c_l^* = 0, \quad (l, j = 1, \ldots, N, \ l, j \neq k1, \ldots, ko). \qquad (27)$$

Equation (27) implies that the system will converge to the set $E_5 = \text{span}\{|\phi_{k1}\rangle, |\phi_{k2}\rangle, \ldots, |\phi_{ko}\rangle\}; l \in \{1, \ldots, N\}, l \neq k1, \ldots, ko\}$, which means that the set $E_5$ contains infinite elements. Therefore this control method cannot ensure the system converge to the target superposition state by adding the restriction defined by (15). But this control method can ensure that the system will converge to the superposition of the eigenstates corresponding to target state. In fact, once one element of $P$ changes, all the populations of the levels will change accordingly. Therefore when there are not very many eigenstates corresponding to target state, the system maybe converge to the target state by regulating the eigenvalues of $P$. Otherwise, the system maybe cannot converge to the target state. Consider the extreme situation in which the target state is the superposition of all the eigenstates. According to the design principle of $P$, we should design $P = c_p I$, where $c_p$ is a real number, and $I$ is the unit matrix. Obviously, in this special case, the control method proposed in Section 2.1.3 will become invalid.

From Section 2.2, one can see that, for the control system governed by (20), $\dot{V}(t) = 0$ holds for all $t \geq t_0, t_0 \in R$ is equivalent to $(H_k)_{jl}(P_l - P_j)\rho_{lj}(t) = 0, (j, l = 1, \ldots, N, \ j < l$. Consider the target state which does not commute with $H_0$, which includes the superposition state and the mixed state which does not commute with $H_0$. Without loss of generality, assume $(\rho_f)_{3N} = (\rho_f)_{N3} \neq 0$; one can design $P_3 = P_N$; thus the system will converge to $E_6 = \{\rho | (\rho)_{ij} = 0, (i, j) \neq (3, N), (i, j) \neq (N, 3), i, j \in \{1, \ldots, N\}\}$. Because the set $E_6$ has infinite elements, this control method also cannot ensure the system will converge to the target state by adding the restriction defined by (26). But when the target state does not have so many nonzero off-diagonal elements, the system maybe converges to the target state by regulating the diagonal elements of $P$. When there are many nonzero off-diagonal elements in the target state, the degree of freedom of $P$ may be not enough.

From the above analyses, we can conclude that by means of using the control methods proposed in Sections 2.1 and 2.2, the control system may but cannot be ensured to converge to the target superposition state. This problem can be solved by means of designing the control laws in the interaction picture of the control system.

Consider the $N$-level control system in the interaction picture as

$$i\frac{\partial}{\partial t}\rho(t) = \left[ \sum_{k=1}^{r} H_k(t) u_k(t), \rho(t) \right]. \qquad (28)$$

Choose the Lyapunov function based on the average value of an imaginary mechanical quantity defined by (21). The first order time derivative of the Lyapunov function (21) can be obtained as

$$\dot{V}_5 = -\sum_{k=1}^{r} u_k(t) \text{tr}\left( iH_k(t) [\rho(t), P] \right). \qquad (29)$$

To ensure $\dot{V}_5 \leq 0$, one can design $u_k(t)$ as

$$u_k(t) = -K_k \operatorname{tr}\left(iH_k(t)\left([\rho(t), P]\right)\right), \qquad K_k > 0, \qquad (30)$$

where $K_k > 0$ and $y_k = f_k(x_k)$, $(k = 1, \ldots, r)$ are monotonic increasing functions through the coordinate origin of plane $x_k - y_k$.

The control system governed by (28) is a non-autonomous system; thus the LaSalle invariance principle cannot be used to analyze the convergence. One can use the improved Barbalat lemma which can be used for the non-autonomous system. According to the improved Barbalat lemma, the convergence of the control system governed by (28) can be depicted by Theorem 7.

**Theorem 7.** *Consider the control system governed by* (28) *with the control field* $u_k(t)$ *designed in* (30). *If (i) the internal Hamiltonian is strongly regular; (ii) the control Hamiltonians are full connected, then any state trajectory will converge to the limit set* $\mathscr{R}_1 = \{\rho_s : [\rho_s, P] = D\}$ *at* $t \to \infty$, *where* $D$ *is a diagonal matrix.*

For the case that $P$ is chosen as a diagonal matrix, the limit set is reduced to $\mathscr{R}_1 = \{\rho_s : [\rho_s, P] = 0\}$. For the case that $P$ is chosen as a nondiagonal matrix, if rank $\widetilde{A}(\vec{P}) = n^2 - n$ holds, then the limit set $\mathscr{R}_1$ is regular; namely, $\mathscr{R}_1 = \{\rho_s : [\rho_s, P] = 0\}$, where $\widetilde{A}(\vec{P})$ is the first $n^2 - n$ rows of $A(\vec{P})$, $A(\vec{P})$ is the real $(n^2 - 1) * (n^2 - 1)$ matrix corresponding to the Bloch representation of $Ad_P$, and $Ad_P$ is a linear map from Hermitian or anti-Hermitian matrices into $su(n)$. $[\rho_s, P] = 0$ means that $P$ and $H_0$ have the same eigenstates. For the target state $\rho_f = |\psi_f\rangle\langle\psi_f|$ being a superposition state, in order to make the target state contain in $\mathscr{R}_1 = \{\rho_s : [\rho_s, P] = 0\}$, $P$ can be designed as

$$P = P_1 |\psi_f\rangle \langle \psi_f| + \sum_{k=2}^{n} P_k |\psi_k\rangle \langle \psi_k|, \qquad (31)$$

where $|\psi_1\rangle = |\psi_f\rangle$ and $\langle\psi_i|\psi_j\rangle = 0$, for $i \neq j$.

Then some deduction shows that $\rho_s = \lambda_i|\psi_i\rangle\langle\psi_i|$($\lambda_i = 1$). In order to make the system converge to the target superposition state $\rho_f$ from any initial pure state, one can design $P$ such that

$$V_5\left(\rho_f\right) < V_5\left(\rho_0\right) < V_5\left(\rho_{\text{oth}}\right), \qquad (32)$$

where $\rho_{\text{oth}}$ represents any other state in the set $\mathscr{R}_1$ except the target state $\rho_f$.

For the satisfaction of $V_5(\rho_f) < V_5(\rho_{\text{oth}})$, one can design $P$ based on Proposition 6. In order to make (32) holds, one must design $P_i$, $(i = 1, \ldots, N)$ such that

$$0 < P_1 < P_1 \operatorname{tr}\left(\rho_f\rho_0\right) + \sum_{k=2}^{N} P_k \operatorname{tr}\left(\rho_k\rho_0\right) < P_j \left(j \neq 1\right). \qquad (33)$$

However, for any initial pure state the target superposition state, there may not exist a $P$ such that (33) holds.

In conclusion, the design principle of $P$ for the case that the target state $\rho_f = |\psi_f\rangle\langle\psi_f|$ being a superposition state is (31), (33), and Proposition 6. We can also conclude that if the control system satisfies the conditions (i) and (ii) in Theorem 7 and there exists a $P$ to make (33) hold, any state trajectory will converge to the target superposition state $\rho_f$ from any initial pure state $\rho_0$.

*Remark 8.* One can also solve the problem of the convergence of the control system governed by (1) to the target superposition state by means of designing the control laws in the interaction picture of the control system.

*2.4. Target Mixed State Which Does Not Commute with the Internal Hamiltonian.* From Sections 2.2 and 2.3, one can see that the control method proposed in Section 2.2 cannot guarantee the convergence to the target state which does not commute with the internal Hamiltonian. For the target state $\rho_f$ being a mixed state which does not commute with the internal Hamiltonian, that is, $[\rho_f, H_0] \neq 0$, Cong, Liu and Yang proposed the control system in the interaction picture and used a unitary transformation to solve the convergence of the control system to the target state. The basic idea is to use a unitary transformation $\widehat{\rho}_f = U_f^\dagger \rho_f U_f$ to make $[\widehat{\rho}_f, H_0] = 0$ hold. Correspondingly, the control system governed by (28) after this unitary transformation will become

$$i\hbar\dot{\widehat{\rho}}(t) = \left[\sum_{k=1}^{r} \widehat{H}_k(t) u_k(t), \widehat{\rho}(t)\right], \qquad (34)$$

where $\widehat{\rho} = U_f^\dagger \rho U_f$, $\widehat{H}_k(t) = U_f^\dagger H_k(t) U_f$.

Then control laws can be designed according to Section 2.3. The research results show that the designed control laws and Theorem 7 are also valid with every physical quantity changing accordingly. In order to make $\widehat{\rho}_f$ contain in $\mathscr{R}_2 \equiv \{\widehat{\rho}_s : [\widehat{\rho}_s, P] = 0\}$, $P$ needs to be designed such that $[H_0, P] = 0$ holds. $[\widehat{\rho}_s, P] = 0$ and $[H_0, P] = 0$ imply that $\widehat{\rho}_s$, $P$, and $H_0$ have the same eigenstates. Since the evolution of $\widehat{\rho}(t)$ is unitary, all the $\widehat{\rho}(t)$ for $t \geq 0$ are isospectral. So the eigenvalues of $\widehat{\rho}_s$ are a permutation of the eigenvalues of $\widehat{\rho}_0$. Thus the limit set $\mathscr{R}_2 \equiv \{\widehat{\rho}_s : [\widehat{\rho}_s, P] = 0\}$ has countable elements. For sake of the convergence from any initial mixed state $\rho_0$ to any target mixed state $\rho_f$ which does not commute with the internal Hamiltonian $H_0$, $P$ needs to be designed such that

$$V_5\left(\widehat{\rho}_f\right) < V_5\left(\widehat{\rho}_0\right) < V_5\left(\widehat{\rho}_{\text{oth}}\right), \qquad (35)$$

where $\widehat{\rho}_{\text{oth}}$ represents any other state in the set $\mathscr{R}_2 \equiv \{\widehat{\rho}_s : [\widehat{\rho}_s, P] = 0\}$ except $\widehat{\rho}_f$.

In such a way, if there exists a $P$ to make (35) hold, any state trajectory of the system will converge to the target mixed state $\rho_f$ which does not commute with the internal Hamiltonian from any initial mixed state $\rho_0$.

Next, let us analyze how to make the restriction defined by (35) hold. Some deductions show that if one designs $P$ based on Proposition 6 with changing $\rho_f$ into $\widehat{\rho}_f$ accordingly, then $V_5(\widehat{\rho}_f) < V_5(\widehat{\rho}_{\text{oth}})$ will hold. Assume $\rho_0 = U\rho_f U^\dagger$. In order to make the whole restriction defined by (35) hold, $P_i$, $(i = 1, \ldots, N)$ needs to be designed such that

$$
\begin{aligned}
(P)_{ll} > & \left( \sum_{i \neq l}^{n} (P)_{ii} \sum_{j \neq k}^{n} \left( (\widehat{\rho}_f)_{jj} - (\widehat{\rho}_f)_{kk} \right) (U_{ij})^2 \right. \\
& + (P)_{ll} \cdot \sum_{j \neq k, j \neq l}^{n} \left( (\widehat{\rho}_f)_{jj} - (\widehat{\rho}_f)_{kk} \right) (U_{lj})^2 \right) \\
& \times \left( (\widehat{\rho}_f)_{kk} - (\widehat{\rho}_f)_{ll} \right)^{-1},
\end{aligned}
\tag{36}
$$

where $U_{ij}$ is the $(i, j)$th element of $U$.

In conclusion, the design principle of $P$ for the case that the target state $\rho_f$ is a mixed state which does not commute with $H_0$ is

(i) $P_l \neq P_m > 0$ for any $l \neq m \in \{1, 2, \ldots, N\}$;

(ii) **Proposition 6** with changing $\rho_f$ into $\widehat{\rho}_f$;

(iii) Equations (36) and (16).

We can conclude from the above analysis that for the case that the target state is a mixed state which does not commute with $H_0$, if the control system is strongly regular and full connected, and one can seek an imaginary mechanical quantity $P$ to make (36) hold, then the control system can converge from any initial mixed state to the target mixed state which does not commute with the internal Hamiltonian.

## 3. Degenerate Cases

The convergence conditions (i) and (ii) proposed in the Theorem 1 through Theorem 7 are relevant to the internal Hamiltonian $H_0$ and the control Hamiltonians $H_k$, $(k = 1, \ldots, r)$. They are system parameters which cannot be changed. And in practice, many actual systems do not satisfy these convergence conditions. These systems are called in the degenerate cases. In order to solve the convergence of the control systems in degenerate cases, the existing method is to introduce a series of implicit function perturbations into the control laws and choose a Lyapunov function which is an implicit function [37–40].

In this section, we also divide the target state into four categories as: target eigenstate, the target mixed state which commute with the internal Hamiltonian, target superposition state, and the target mixed state which does not commute with the internal Hamiltonian. For these four target state categories, respectively, research results for the degenerate cases are summarized and analyzed.

*3.1. Target Eigenstate.* In this Section, the convergence of the control system based on the Lyapunov control method in the degenerate cases to any target eigenstate from any initial pure state will be summarized and analyzed.

*3.1.1. Implicit Lyapunov Control Based on State Distance.* In order to solve the convergence of the control systems in the degenerate cases, several researchers introduced a series of implicit function perturbations into the control laws [37–40]. After a series of perturbations $\gamma_k(t)$ introducing into the control laws, the dynamical equation (1) becomes

$$
i \, | \dot{\psi}(t) \rangle = \left( H_0 + \sum_{k=1}^{r} H_k \left( \gamma_k(t) + v_k(t) \right) \right) | \psi(t) \rangle, \tag{37}
$$

where $\gamma_k(t) + v_k(t) = u_k(t)$, $(k = 1, \ldots, r)$ are the total control laws.

Without loss of generality, assume $|\psi_f\rangle = |\phi_g\rangle$, $1 \leq g \leq N$. In order to solve the problem of convergence for the degenerate cases, the perturbations $\gamma_k(t)$ were introduced into the control laws. The basic idea is as follows: Denote the system with the internal Hamiltonian $H_0$, the control Hamiltonians $H_k$, $(k = 1, \ldots, r)$, and the control laws $u_k(t) = \gamma_k(t) + v_k(t)$ as system 1, and the system with the internal Hamiltonian $H_{02} = H_0 + \sum_{k=1}^{r} H_k \gamma_k(t)$, the control Hamiltonians $H_k$, $(k = 1, \ldots, r)$, and the control laws $v_k(t)$ as system 2. All these two systems can be depicted by (37). Denote the eigenvalues and eigenstates of $H_{02}$ as $\lambda_{1, \gamma_1(t), \ldots, \gamma_r(t)}, \lambda_{2, \gamma_1(t), \ldots, \gamma_r(t)}, \ldots, \lambda_{N, \gamma_1(t), \ldots, \gamma_r(t)}$ and $|\phi_{1, \gamma_1(t), \ldots, \gamma_r(t)}\rangle, |\phi_{2, \gamma_1(t), \ldots, \gamma_r(t)}\rangle, \ldots, |\phi_{N, \gamma_1(t), \ldots, \gamma_r(t)}\rangle$, respectively, which are the functions of the perturbations $\gamma_k(t)$. Assume $|\psi_{f, \gamma_1(t), \ldots, \gamma_r(t)}\rangle = |\phi_{g, \gamma_1(t), \ldots, \gamma_r(t)}\rangle$. If one can design the perturbations $\gamma_k(t)$ such that (i) $\omega_{l, m, \gamma_1, \ldots, \gamma_r} \neq \omega_{i, j, \gamma_1, \ldots, \gamma_r}$, $(l, m) \neq (i, j), i, j, l, m \in \{1, 2, \ldots, N\}$, where $\omega_{l, m, \gamma_1, \ldots, \gamma_r} = \lambda_{l, \gamma_1, \ldots, \gamma_r} - \lambda_{m, \gamma_1, \ldots, \gamma_r}$ holds; (ii) for any $|\phi_{\gamma_1, \ldots, \gamma_r}\rangle \neq |\psi_{f, \gamma_1, \ldots, \gamma_r}\rangle$, there exists at least a $k \in \{1, \ldots, r\}$ satisfying $\langle \phi_{\gamma_1, \ldots, \gamma_r} | H_k | \psi_{f, \gamma_1, \ldots, \gamma_r}\rangle \neq 0$, and select the specific Lyapunov function based on the state distance as

$$
V_6(t) = \frac{1}{2} \left( 1 - \left| \langle \psi \mid \psi_{f, \gamma_1(t), \ldots, \gamma_r(t)} \rangle \right|^2 \right). \tag{38}
$$

Then according to Section 2.1.1, system 2 will converge to $|\psi_{f, \gamma_1(t), \ldots, \gamma_r(t)}\rangle$. And when system 2 converge to $|\psi_{f, \gamma_1(t), \ldots, \gamma_r(t)}\rangle$, if the perturbations $\gamma_k(t)$ at $|\psi_{f, \gamma_1(t), \ldots, \gamma_r(t)}\rangle$ are designed to equal zero, then system 2 will become system 1, and $|\psi_{f, \gamma_1(t), \ldots, \gamma_r(t)}\rangle$ will become $|\psi_f\rangle$. Then the convergence of system 1 to $|\psi_f\rangle$ will be ensured. In fact, the evolution of system 1 can be viewed as a composite of two evolution processes. One is system 2 converges to $|\psi_{f, \gamma_1(t), \ldots, \gamma_r(t)}\rangle$ from the initial state $|\psi_0\rangle$, another one is $\gamma_k(t)$ converges to 0. In order to make system 1 in the non-degenerate case converge to $|\psi_f\rangle$, the speed of $\gamma_k(t)$, $(k = 1, \ldots, r)$ converging to 0 must be slower than the speed at which system 2 converges toward $|\psi_{f, \gamma_1(t), \ldots, \gamma_r(t)}\rangle$. For convenience, the control system in the following section means system 1.

The existing design method of $\gamma_k(t)$ is to design it to be a monotonically increasing functional of $V(t)$, that is,

$$
\gamma_k(t) = \theta_k(V(t)) = \theta_k \left( \frac{1}{2} \left( 1 - \left| \langle \psi \mid \psi_{f, \gamma_1(t), \ldots, \gamma_r(t)} \rangle \right|^2 \right) \right),
$$
$$
(k = 1, \ldots, r), \tag{39}
$$

where the function $\theta_k(\cdot)$ satisfies $\theta_k(0) = 0$, $\theta_k(s) > 0$ and $\theta_k'(s) > 0$ for every $s > 0$ and $s$ is the independent variable of the function $\theta_k(\cdot)$.

The right side of (39) contains $|\psi_{f,\gamma_1(t),\ldots,\gamma_r(t)}\rangle$ which is a functional of the perturbations $\gamma_k(t),(k = 1,\ldots,r)$. One can see that the relation between $\gamma_k$ and the time $t$ is defined by $r$ equations and cannot be expressed by an explicit expression, so $\gamma_k$ is the implicit function of the time $t$. From (38) and (39), one can also see that the Lyapunov function $V_6$ is the implicit function of the time $t$. The existence of $\gamma_k(t)$ can be depicted by Lemma 9.

**Lemma 9** (see [37, 40]). *Let* $\theta_k \in C^\infty(R^+; [0, \gamma_k^*]), k = 1,\ldots,r, (\gamma_k^* \in R, \gamma_k^* > 0)$ *satisfy* $\theta_k(0) = 0,\ \theta_k(s) > 0,\ \theta_k'(s) > 0$ *for every* $s > 0,\ \|\theta_k'\|_\infty < 1/C^*$, *and* $C^* = 1 + \max\{\|(\partial\psi_{f,\gamma_1(t),\ldots,\gamma_r(t)}/\partial\gamma_k)\|_{(\gamma_{10},\ldots,\gamma_{r0})}\|; \gamma_{k0} \in [0, \gamma_k^*]\}$, *where* $\gamma_{k0} = \gamma_k(0)$. *Then for every state* $|\psi\rangle \in S^{2N-1} = \{x \in C^N; \|x\| = 1\}$, *there exists a unique* $\gamma_1, \gamma_2,\ldots$, *and* $\gamma_r$ *with* $\gamma_k \in C^\infty(\gamma_k \in [0, \gamma_k^*]), (k = 1,\ldots,r)$ *satisfying* $\gamma_k(t) = \theta_k((1/2)(1 - |\langle\psi \mid \psi_{f,\gamma_1(t),\ldots,\gamma_r(t)}\rangle|^2)), (k = 1,\ldots,r), \gamma_k(|\psi_f\rangle) = 0$.

By (38) and **Lemma 9**, one can see that for every state $|\psi\rangle \in S^{2N-1} = \{x \in C^N; \|x\| = 1\}$; there also exists a unique $V$.

Assume

$$\frac{\partial |\psi_{f,\gamma_1(t),\ldots,\gamma_r(t)}\rangle}{\partial\gamma_1} = \frac{\partial |\psi_{f,\gamma_1(t),\ldots,\gamma_r(t)}\rangle}{\partial\gamma_2} = \cdots = \frac{\partial |\psi_{f,\gamma_1(t),\ldots,\gamma_r(t)}\rangle}{\partial\gamma_r}. \tag{40}$$

The first order time derivative of the Lyapunov function (38) is

$$\dot{V}_6 = \frac{-\sum_{k=1}^r v_k(t)\,\mathfrak{I}\left(\langle\psi \mid \psi_{f,\gamma_1,\ldots,\gamma_r}\rangle\langle\psi_{f,\gamma_1,\ldots,\gamma_r} \mid H_k \mid \psi\rangle\right)}{(1 + \mathfrak{R}\left(\langle\psi \mid \psi_{f,\gamma_1,\ldots,\gamma_r}\rangle\langle(\partial \mid \psi_{f,\gamma_1,\ldots,\gamma_r}\rangle/\partial\gamma_k) \mid \psi\rangle\right)\sum_{j=1}^r \theta_j')}. \tag{41}$$

In order to make $\dot{V}_6(t) \leq 0$, one can design $\gamma_k(t)$ such that $\|\theta_j'\|_\infty < 1/(rC^*)$ to make $1/(1 + \mathfrak{R}(\langle\psi \mid \phi_{\gamma_1,\ldots,\gamma_r}\rangle\langle(\partial\phi_{\gamma_1,\ldots,\gamma_r}/\partial\gamma_k) \mid \psi\rangle)\sum_{j=1}^r \theta_j') > 0$ hold and design $v_k(t)$ as

$$v_k(t) = K_k f_k\left(\mathfrak{I}\left(e^{i\angle\langle\psi|\psi_{f,\gamma_1,\ldots,\gamma_r}\rangle}\langle\psi_{f,\gamma_1,\ldots,\gamma_r}| H_k|\psi\rangle\right)\right), \tag{42}$$

where $K_k > 0$, and $v_k(t), (k = 1,\ldots,r)$ are monotonic functions through the coordinate origin and in the first quadrant and the third quadrant.

Based on the LaSalle invariance principle, the convergence of the control system governed by (37) can be depicted by Theorem 10.

**Theorem 10** (see [37, 40]). *Consider the control system* (37) *with control fields* $u_k(t) = \gamma_k(t) + v_k(t), \gamma_k(t)$ *designed in* (39), *Lemma 9 and* $\|\theta_j'\|_\infty < 1/(rC^*)$, *and* $v_k(t)$ *designed in* (42). *Assume that the target state* $|\psi_f\rangle$ *is an eigenstate of* $H_0$. *If the control system satisfies* (i) $\omega_{l,m,\gamma_1,\ldots,\gamma_r} \neq \omega_{i,j,\gamma_1,\ldots,\gamma_r}, (l,m) \neq (i,j)$, $i,j,l,m \in \{1,2,\ldots,N\}, \omega_{l,m,\gamma_1,\ldots,\gamma_r} = \lambda_{l,\gamma_1,\ldots,\gamma_r} - \lambda_{m,\gamma_1,\ldots,\gamma_r}$, *where* $\lambda_{l,\gamma_1,\ldots,\gamma_r}$ *is an eigenvalue of* $(H_0 + \sum_{k=1}^r H_k\gamma_k)$, (ii) *For any* $|\phi_{\gamma_1,\ldots,\gamma_r}\rangle \neq |\psi_{f,\gamma_1,\ldots,\gamma_r}\rangle$, *there exists at least a* $k \in \{1,\ldots,r\}$

*satisfying* $\langle\phi_{\gamma_1,\ldots,\gamma_r}|H_k|\psi_{f,\gamma_1,\ldots,\gamma_r}\rangle \neq 0$; *then the largest invariant set is* $S^{2N-1} \cap E_7$, *where* $E_7 = \{|\psi\rangle = e^{i\theta}|\psi_f\rangle, \theta \in R\}$. *And the control system will converge toward* $|\psi_f\rangle e^{i\theta}, (\theta \in R)$.

Conditions (i) and (ii) in **Theorem 10** are associated with $H_0, H_k, (k = 1,\ldots,r)$ and $\gamma_k(t)$. By designing appropriate $\gamma_k(t)$, these two conditions can be satisfied in most cases. one can see that if one designs appropriate control laws $u_k(t) = \gamma_k(t) + v_k(t), (k = 1,\ldots,r)$ to make the conditions (i) and (ii) hold, the control system depicted by (37) can converge to the equivalent state of the target eigenstate $|\psi_f\rangle$ from any initial pure state.

### 3.1.2. Implicit Lyapunov Control Based on the State Error.
The basic idea of this method is similar to that of the state distance. In the Lyapunov control based on the state error, a global phase control item $\omega$ is added into the control system to facilitate the design of the control laws. Thus the control system can be depicted by

$$i\,|\dot{\psi}(t)\rangle = \left(H_0 + \sum_{k=1}^r H_k\left(\gamma_k(t) + v_k(t)\right) + \omega I\right)|\psi(t)\rangle, \tag{43}$$

where $u_k(t) = \gamma_k(t) + v_k(t), (k = 1,\ldots,r)$ and $\omega$ is the control laws which need to design.

Consider the Lyapunov function based on the state error as

$$V_7(t) = \frac{1}{2}\left\langle\psi - \psi_{f,\gamma_1,\ldots,\gamma_r} \mid \psi - \psi_{f,\gamma_1,\ldots,\gamma_r}\right\rangle. \tag{44}$$

Thus the implicit function perturbation $\gamma_k(t)$ needs to be designed as

$$\gamma_k(t) = \theta_k(V(t)) = \theta_k\left(\frac{1}{2}\left\langle\psi - \psi_{f,\gamma_1,\ldots,\gamma_r} \mid \psi - \psi_{f,\gamma_1,\ldots,\gamma_r}\right\rangle\right), \tag{45}$$

where functions $\theta_k(\cdot)$ satisfy $\theta_k(0) = 0, \theta_k(s) > 0, \theta_k'(s) > 0$ for every $s > 0$.

The existence of $\gamma_k(t)$ can be depicted by Lemma 11.

**Lemma 11** (see [38, 39]). *Let* $\theta_k \in C^\infty(R^+; [0, \gamma_k^*]), k = 1,\ldots,r, (\gamma_k^* \in R, \gamma_k^* > 0)$ *satisfy* $\theta_k(0) = 0, \theta_k(s) > 0, \theta_k'(s) > 0$ *for every* $s > 0, \|\theta_k'\|_\infty < 1/rC^*$, *and* $C^* = 1 + \max\{\|(\partial|\psi_{f,\gamma_1,\ldots,\gamma_r}\rangle/\partial\gamma_k)\|_{(\gamma_{10},\ldots,\gamma_{r0})}\|; \gamma_{k0} \in [0, \gamma_k^*]\}$. *Then for every state* $|\psi\rangle \in S^{2N-1} = \{x \in C^N; \|x\| = 1\}$, *there exists a unique* $\gamma_1, \gamma_2,\ldots$, *and* $\gamma_r$ *with* $\gamma_k \in C^\infty(\gamma_k \in [0, \gamma_k^*]), (k = 1,\ldots,r)$ *satisfying* $\gamma_k(t)) = \theta_k((1/2)\langle\psi - \psi_{f,\gamma_1(|\psi\rangle),\ldots,\gamma_r(|\psi\rangle)} \mid \psi - \psi_{f,\gamma_1(|\psi\rangle),\ldots,\gamma_r(|\psi\rangle)}\rangle)), \gamma_k(|\psi_f\rangle) = 0$.

Assume (40) holds, the first order time derivative of the Lyapunov function (44) is

$$
\dot{V}_7 = -\left( \frac{1}{1 + \Re\left( \left\langle \left( \partial \left| \psi_{f,\gamma_1,\dots,\gamma_r} \right\rangle / \partial\gamma_k \right) \middle| \psi \right\rangle \sum_{j=1}^{r} \theta'_j \right)} \right)
$$
$$
\cdot \left( \left( \lambda_{\gamma_1,\dots,\gamma_r} + \omega \right) \Im\left( \psi_{f,\gamma_1,\dots,\gamma_r} \mid \psi \right) \right.
$$
$$
\left. + \sum_{k=1}^{r} \Im\left( \left\langle \psi_{f,\gamma_1,\dots,\gamma_r} \middle| H_k \middle| \psi \right\rangle \right) v_k(t) \right).
$$
(46)

In order to make $\dot{V}_7(t) \leq 0$, $\omega$ and $v_k(t)$ need to be designed as

$$
\omega = -\lambda_{\gamma_1,\dots,\gamma_r} + c f_0 \left( \Im\left( \left\langle \psi_{f,\gamma_1,\dots,\gamma_r} \mid \psi \right\rangle \right) \right), \qquad (47)
$$

$$
v_k(t) = K_k f_k \left( \Im\left( \left\langle \psi_{f,\gamma_1,\dots,\gamma_r} \middle| H_k \middle| \psi \right\rangle \right) \right), \quad (k = 1,\dots,r), \qquad (48)
$$

where $K_k > 0$, $c > 0$, and $y_k = f_k(x_k), (k = 0, 1, \dots, r)$ are monotonic functions through the coordinate origin and in the first quadrant and the third quadrant.

Based on the LaSalle invariance principle, the convergence of the control system governed by (44) can be depicted by Theorem 12.

**Theorem 12** (see [38, 39]). *Consider the control system (43) with control fields $\gamma_k(t)$ designed in (45) and Lemma 11, $v_k(t)$ designed in (48), and $\omega$ designed in (47). Assume that the target state $|\psi_f\rangle$ is an eigenstate of $H_0$. If the control system satisfies (i) $\omega_{l,m,\gamma_1,\dots,\gamma_r} \neq \omega_{i,j,\gamma_1,\dots,\gamma_r}, (l,m) \neq (i,j), i, j, l, m \in \{1,\dots,N\}, \omega_{l,m,\gamma_1,\dots,\gamma_r} = \lambda_{l,\gamma_1,\dots,\gamma_r} - \lambda_{m,\gamma_1,\dots,\gamma_r}, \lambda_{l,\gamma_1,\dots,\gamma_r}$ is the lth eigenvalue of $(H_0 + \sum_{k=1}^{r} H_k(\gamma_k + \eta_k))$ corresponding to the eigenstate $|\phi_{l,\gamma_1,\dots,\gamma_r}\rangle$; (ii) There is at least a $k \in \{1,\dots,r\}, \langle \phi_{\gamma_1(t),\dots,\gamma_r(t)} | H_k | \psi_{f,\gamma_1(t),\dots,\gamma_r(t)} \rangle \neq 0$, for $|\phi_{\gamma_1(t),\dots,\gamma_r(t)}\rangle \neq |\psi_{f,\gamma_1(t),\dots,\gamma_r(t)}\rangle$; then the largest invariant set is $S^{2N-1} \cap E_8$, where $E_8 = \{|\psi\rangle = e^{i\theta}|\psi_f\rangle, \theta \in R\}$. And the control system will converge toward the equivalent state of the target state: $|\psi_f\rangle e^{i\theta}, (\theta \in R)$.*

By designing appropriate perturbations $\gamma_k(t)$, conditions (i) and (ii) in Theorem 12 can be satisfied in most cases. From Theorem 12, one can see that if one design appropriate control laws $u_k(t) = \gamma_k(t) + v_k(t), (k = 1,\dots,r)$ and $\omega$ to make the conditions (i) and (ii) in Theorem 12 hold, the control system depicted by (43) can converge to the equivalent state of the target eigenstate $|\psi_f\rangle$ from any initial pure state.

### 3.1.3. Implicit Lyapunov Control Based on Average Value of an Imaginary Mechanical Quantity.

For the implicit Lyapunov control based on the average value of an imaginary mechanical quantity, one can consider the control system depicted by (37). In this control method, the introduced implicit function perturbations $\gamma_k(t)$ mainly have two tasks. One task is to solve the convergent problem of the control system in the degenerate cases. The basic idea is similar to that of the state distance. Another one is to choose a simpler restriction $V(|\psi_f\rangle) < V(|\psi_{\text{other}}\rangle)$, which can be satisfied for any initial state and any target state by designing the imaginary mechanical quantity. In order to ensure the system converge to the target state by adding $V(|\psi_f\rangle) < V(|\psi_{\text{other}}\rangle)$, we can design all the perturbations $\gamma_k(t) = 0$ hold for $k = 1,\dots,r$ only at $|\psi_f\rangle$, that is, (1) $\gamma_k(|\psi_f\rangle) = 0, (k = 1,\dots,r)$, and (2) for $|\psi\rangle \neq |\psi_f\rangle$, there exists at least one $k$ such that $\gamma_k(|\psi\rangle) \neq 0$. For sake of completing these two tasks, we can design $\gamma_k(t)$ as a monotonically increasing functional of $V(t)$:

$$
\gamma_k(|\psi\rangle) = C_k \cdot \theta_k \left( V(|\psi\rangle) - V(|\psi_f\rangle) \right), \qquad (49)
$$

where $C_k \geq 0$, and for $k = 1,\dots,r$, there exists at least a $C_k > 0$, and the function $\theta_k(\cdot)$ satisfies $\theta_k(0) = 0, \theta_k(s) > 0$ and $\theta'_k(s) > 0$ for every $s > 0$, $s$ is the independent variable of the function $\theta_k(\cdot)$.

The specific Lyapunov function based on the average value of an imaginary mechanical quantity can be selected as:

$$
V_8(|\psi\rangle) = \langle \psi| P_{\gamma_1,\dots,\gamma_r} |\psi\rangle, \qquad (50)
$$

where $P_{\gamma_1,\dots,\gamma_r} = f(\gamma_1(t),\dots,\gamma_r(t))$ is a functional of $\gamma_k(t)$ and positive definite.

The existence of $\gamma_k(t)$ can be depicted by Lemma 13.

**Lemma 13.** *If $C_k = 0, \gamma_k(|\psi\rangle) = 0$. Else if $C_k > 0, \theta_k \in C^\infty(R^+; [0, \gamma_k^*]), k = 1,\dots,r$ ($\gamma_k^*$ is a positive constant) satisfy $\theta_k(0) = 0, \theta_k(s) > 0$ and $\theta'_k(s) > 0$ for every $s > 0$, and $|\theta'_k| < 1/(2C_kC^*), C^* = 1+C, C = \max\{\|\partial P_{\gamma_1,\dots,\gamma_r}/\partial\gamma_k\|_\infty, \gamma_k \in [0, \gamma_k^*]\}$, then for every $|\psi\rangle \in S^{2N-1}$, there is a unique $\gamma_k \in C^\infty(\gamma_k \in [0, \gamma_k^*])$ satisfying $\gamma_k(|\psi\rangle) = C_k \cdot \theta_k(\langle\psi|P_{\gamma_1,\dots,\gamma_r}|\psi\rangle - \langle\psi_f|P_{\gamma_1,\dots,\gamma_r}|\psi_f\rangle)$.*

*For the sake of simplicity, $\gamma_k(t)$ can be designed zero for some k, and other $\gamma_k(t)$ equal, denoted by $\gamma(t)$; that is, set*

$$
\gamma_k(t) = \gamma(t) = \theta\left( \langle \psi| P_\gamma |\psi\rangle - \langle \psi_f| P_\gamma |\psi_f\rangle \right),
$$
$$
k = k_1,\dots,k_m; \qquad (51)
$$
$$
\gamma_k(t) = 0, \quad k \neq k_1,\dots,k_m \ (1 \leq k_1,\dots,k_m \leq r),
$$

*where $\theta(\cdot) = \theta_{k_1}(\cdot) = \cdots = \theta_{k_m}(\cdot)$ and $P_\gamma$ are functionals of $\gamma(t)$.*

The time derivative of the selected Lyapunov function is

$$
\dot{V}_8 = \sum_{k=1}^{r} i v_k(t) \langle \psi| \left[ H_k, P_\gamma \right] |\psi\rangle
$$
$$
\cdot \frac{\left( 1 + \theta'\left( \langle \psi_f| \left( \partial P_\gamma/\partial\gamma \right) |\psi_f\rangle \right) \right)}{\left( 1 - \theta'\left( \langle \psi| \left( \partial P_\gamma/\partial\gamma \right) |\psi\rangle - \langle \psi_f| \left( \partial P_\gamma/\partial\gamma \right) |\psi_f\rangle \right) \right)}. \qquad (52)
$$

In order to ensure $\dot{V}_8 \leq 0$, $v_k(t)$ can be designed as

$$
v_k(t) = -K_k f_k \left( i \langle \psi| \left[ H_k, P_\gamma \right] |\psi\rangle \right), \quad (k = 1,\dots,r), \qquad (53)
$$

where $K_k$ is a constant and $K_k > 0$, and $y_k = f_k(x_k), (k = 1, 2, \ldots, r)$ are monotonic increasing functions through the coordinate origin of the plane $x_k - y_k$.

Based on LaSalle's invariance principle, the convergence of the control system governed by (37) can be depicted by Theorem 14.

**Theorem 14.** *Consider the control system governed by (37) with control fields $u_k(t) = \gamma_k(t) + v_k(t), (k = 1, \ldots, r)$, where $\gamma_k(t)$ is defined by Lemma 13 and (51) and $v_k(t)$ is defined by (53). If the control system satisfies (i) $\omega_{l,m,\gamma} \neq \omega_{i,j,\gamma}, (l, m) \neq (i, j)$, $i, j, l, m \in \{1, 2, \ldots, N\}$, $\omega_{i,m,\gamma} = \lambda_{l,\gamma} - \lambda_{m,\gamma}$, where $\lambda_{l,\gamma}$ is the $l$th eigenvalue of $H_0 + \sum_{n=k_1}^{k_m} H_n \gamma(t)$ corresponding to the eigenstate $|\phi_{l,\gamma}\rangle$; (ii) for any $i \neq j$, $i, j \in \{1, 2, \ldots, N\}$, there exists at least one $k$ such that $(\widehat{H}_k)_{lm} \neq 0$, where $(\widehat{H}_k)_{lm}$ is the $(l, m)$th element of $\widehat{H}_k = U_1^\dagger H_k U_1$, $U_1 = (|\phi_{1,\gamma}\rangle, \ldots, |\phi_{N,\gamma}\rangle)$; (iii) $[P_\gamma, H_0 + \sum_{n=k_1}^{k_m} H_n \gamma(t)] = 0$; (iv) $(\widehat{P}_\gamma)_{ll} \neq (\widehat{P}_\gamma)_{mm}, l \neq m$, where $(\widehat{P}_\gamma)_{ll}$ is the $(l, l)$th element of $U_1^H P_\gamma U_1$, then any trajectory will converge toward $E_9 = \{|\psi_{l,\gamma}\rangle|e^{i\theta_l}|\phi_{l,\gamma(|\psi_{l,\gamma}\rangle)}\rangle; \theta_l \in R, l \in \{1, \ldots, N\}\}$.*

From Theorem 14, one can see that if the target state $|\psi_f\rangle$ is an eigenstate, $|\psi_f\rangle$ is contained in $E_9$. In order to make the system converge to the target eigenstate from any initial pure state $|\psi_0\rangle$, on the one hand, $P_\gamma$ needs to be designed to make

$$V_8\left(|\psi_f\rangle\right) < V_8\left(|\psi_{other}\rangle\right) \tag{54}$$

hold, where $|\psi_{other}\rangle$ represents any other state in the set $E_9$ except the target state. On the other hand, because $\partial \gamma/\partial V > 0, \dot{V} \leq 0, \gamma \geq 0$ holds, when $v_k(t) = 0, \gamma(t) = \overline{\gamma} \neq 0$ holds for some time, we can design $\gamma = \overline{\gamma} - \alpha, (0 < \alpha \ll \overline{\gamma})$ to make the state trajectory evolve but not stay in $E_9$ until $|\psi_f\rangle e^{i\theta_l}$ is reached.

Next, let us analyze how to realize the convergence conditions in Theorem 14 and restriction defined by (54). By designing appropriate $\gamma_k(t)$, conditions (i) and (ii) in Theorem 14 can be satisfied in most cases. In order to make condition (iii) be satisfied, the eigenvalues of $P_\gamma$ can be designed constant, denoted by $P_1, P_2, \ldots, P_N$, and $\widehat{P}_\gamma$ can be designed as

$$P_\gamma = \sum_{j=1}^{N} P_j |\phi_{j,\gamma}\rangle \langle \phi_{j,\gamma}|. \tag{55}$$

To make condition (iv) satisfied, $P_l \neq P_j (\forall l \neq j; 1 \leq l, j \leq N)$ can be designed. The results on how to make (54) hold can be depicted by Theorem 15.

**Theorem 15.** *If one designs $P_i > P_f, (i = 1, \ldots, N, P_i \neq P_f)$, then $V_8(|\psi_f\rangle) < V_8(|\psi_{other}\rangle)$ holds, where $P_f$ is the eigenvalue of $P_{\gamma(|\psi_f\rangle)}$ corresponding to $|\psi_f\rangle$.*

We can conclude from the above analyses that by using the implicit Lyapunov control based on the imaginary mechanical quantity, if one designs appropriate control laws

$u_k(t) = \gamma_k(t) + v_k(t), (k = 1, \ldots, r)$ to make the conditions (i)–(iv) in Theorem 15 and (54) hold, the control system governed by (37) in the degenerate cases can converge from any initial pure state to the target eigenstate. The design principle of the imaginary mechanical quantity $P_\gamma$ is Theorem 15 and (55).

### 3.2. Target Superposition State.

The method proposed in Section 3.1 cannot guarantee the control system governed by (37) or (43) in the degenerate cases converge from any initial pure state to the target superposition state. In order to solve this problem, one can introduce a series of constant disturbances $\eta_k$ into the control laws. Thus the mechanical equation (37) becomes

$$i |\dot{\psi}(t)\rangle = \left(H_0 + \sum_{k=1}^{r} H_k \left(\eta_k + \gamma_k(t) + v_k(t)\right)\right)|\psi(t)\rangle. \tag{56}$$

And the mechanical equation (43) in Section 3.1.2 becomes

$$i |\dot{\psi}(t)\rangle = \left(H_0 + \sum_{k=1}^{r} H_k \left(\gamma_k(t) + v_k(t) + \eta_k\right) + \omega I\right)|\psi(t)\rangle, \tag{57}$$

where $\eta_k \in R$.

The basic idea of solving the convergence to the target superposition state is to design $\eta_k$ to make the target state $|\psi_f\rangle$ be an eigenstate of $H_0' = H_0 + \sum_{k=1}^{r} H_k \eta_k$. $H_0'$ can be viewed as the new internal Hamiltonian of the control system. If the number of the control Hamiltonians $r$ is large enough, by designing appropriate $\eta_k$, $(H_0 + \sum_{k=1}^{r} H_k \eta_k)|\psi_f\rangle = \lambda_f'|\psi_f\rangle$ can be satisfied in most cases, where $\lambda_f'$ is the eigenvalue of $H_0 + \sum_{k=1}^{r} H_k \eta_k$ corresponding to $|\psi_f\rangle$. Then one can design the control laws and analyze the convergence according to the method for the target eigen state cases. Research results show that every conclusion in Section 3.1 also holds with changing $H_0$ into $H_0'$.

### 3.3. Target Mixed State Which Commutes with the Internal Hamiltonian.

Consider the $N$-level closed quantum system governed by the following quantum Liouville equation:

$$i \dot{\rho}(t) = \left[H_0 + \sum_{k=1}^{r} H_k \left(\gamma_k(t) + v_k(t)\right), \rho(t)\right], \tag{58}$$

where $\gamma_k(t) + v_k(t) = u_k(t)$ are the total control laws.

The design ideas are similar to those of Section 3.1. Consider the Lyapunov function based on the imaginary mechanical quantity as

$$V_9(\rho) = \text{tr}\left(P_{\gamma_1, \ldots, \gamma_r}\rho\right), \tag{59}$$

$\gamma_k(t)$ can be designed as

$$\gamma_k(t) = \gamma(t) = \theta\left(V_9(\rho) - V_9(\rho_f)\right), \quad k = k_1, \ldots, k_m;$$

$$\gamma_k(t) = 0, \quad k \neq k_1, \ldots, k_m (1 \leq k_1, \ldots, k_m \leq r). \tag{60}$$

The existence of $\gamma(t)$ can be depicted by Lemma 16.

**Lemma 16.** *If* $\theta \in C^\infty(R^+; [0, \gamma^*])$, $k = 1, \ldots, r$ ($\gamma^*$ *is a positive constant*) *satisfy* $\theta(0) = 0$, $\theta(s) > 0$ *and* $\theta'(s) > 0$ *for every* $s > 0$, *and* $|\theta'| < 1/(2C^*)$, $C^* = 1 + C$, $C = \max\{\|\partial P_\gamma/\partial \gamma\|_{m_1}, \gamma \in [0, \gamma^*]\}$, *then for every* $\rho$, *there is a unique* $\gamma \in C^\infty(\gamma \in [0, \gamma^*])$ *satisfying* $\gamma(\rho) = \theta(\text{tr}(P_\gamma \rho) - \text{tr}(P_\gamma \rho_f))$.

According to the Lyapunov stabilty Theorem, $\nu_k(t)$ can be designed as

$$\nu_k(t) = K_k f_k\left(i\, \text{tr}\left(\left[P_\gamma, H_k\right]\rho\right)\right), \quad (k = 1, \ldots, r), \quad (61)$$

where $K_k$ is a constant and $K_k > 0$, and $y_k = f_k(x_k)$, $(k = 1, 2, \ldots, r)$ are monotonic increasing functions through the coordinate origin of the plane $x_k - y_k$.

Based on the LaSalle invariance principle, the convergence of the control system can be depicted by Theorem 17.

**Theorem 17.** *Consider the control system depicted by* (58) *with control laws* $u_k(t) = \gamma_k(t) + \nu_k(t)$, *where* $\gamma_k(t)$ *is defined by Lemma 16 and* (60), *and* $\nu_k(t)$ *is defined by* (61). *If the control system satisfies* (i) $\omega_{l,m,\gamma} \neq \omega_{i,j,\gamma}$, $(l, m) \neq (i, j)$, $i, j, l, m \in \{1, 2, \ldots, N\}$, $\omega_{l,m,\gamma} = \lambda_{l,\gamma} - \lambda_{m,\gamma}$, *where* $\lambda_{l,\gamma}$ *is the lth eigenvalue of* $(H_0 + \sum_{n=k_1}^{k_m} H_n \gamma(t))$ *corresponding to the eigenstate* $|\phi_{l,\gamma}\rangle$; (ii) $\forall j \neq l$, *for* $k = 1, \ldots, r$, *there exists at least a* $(\widehat{H}_k)_{jl} \neq 0$, *where* $(\widehat{H}_k)_{jl}$ *is the* $(j,l)$*th element of* $\widehat{H}_k = U_2^H H_k U_2$ *with* $U_2 = (|\phi_{1,\gamma}\rangle, \ldots, |\phi_{N,\gamma}\rangle)$; (iii) $[P_\gamma, H_0 + \sum_{n=k_1}^{k_m} H_n \gamma(t)] = 0$, $1 \leq k_1, \ldots, k_m \leq r$; (iv) *for any* $l \neq j$, $(1 \leq l, j \leq N)$, $(\widehat{P}_\gamma)_{ll} \neq (\widehat{P}_\gamma)_{jj}$ *holds, where* $(\widehat{P}_\gamma)_{ll}$ *is the* $(l, l)$*th element of* $\widehat{P}_\gamma = U_2^H P_\gamma U_2$, *then the control system will converge toward* $E_{10} = \{\rho_{t_0}|(U_2^\dagger \rho_{t_0} U_2)_{ij} = 0, \gamma = \gamma(\rho_{t_0}), t_0 \in R\}$.

For the target state $\rho_f$ which commutes with the internal Hamiltonian; that is, $[\rho_f, H_0] = 0$, $\rho_f$ is contained in $E_{10}$. In order to make the system converge to the target state $\rho_f$ which commutes with $H_0$ from any initial stat; on the one hand, $P_\gamma$ needs to be designed to make

$$V_9\left(\rho_f\right) < V_9\left(\rho_{\text{other}}\right) \quad (62)$$

hold, where $\rho_{\text{other}}$ represents any other state in the set $E_{10}$ except the target state. On the other hand, $\gamma$ was designed as $\gamma = \overline{\gamma} - \alpha$, $(0 < \alpha \ll \overline{\gamma})$ when $\nu_k(t) = 0$, $\gamma(t) = \overline{\gamma} \neq 0$ holds for some time to make the state trajectory evolve but not stay in $E_{10}$ until $\rho_f$ is reached.

For the satisfaction of conditions (i)–(iv) in Theorem 17, one can follow that of Section 3.1.3. In order to make (62) hold, one can design $P_\gamma$ based on Proposition 6. The total design principle of $P_\gamma$ is (55) and Proposition 6.

We can conclude from the above analyses that for the target state which commutes with the internal Hamiltonian, by using the implicit Lyapunov control method based on the imaginary mechanical quantity, if one designs appropriate control laws $u_k(t) = \gamma_k(t) + \nu_k(t)$, $(k = 1, \ldots, r)$ to make the conditions (i)–(iv) in Theorem 17 and (62) hold, the control system governed by (58) in the degenerate cases can converge from any initial state to the target state which commutes with the internal Hamiltonian, which contains the target eigenstate and the target mixed state which commutes with the internal Hamiltonian.

*3.4. Target Mixed State Which Does Not Commute with Internal Hamiltonian.* For the target mixed state which does not commute with the internal Hamiltonian, the design idea is similar to Section 3.2. The difference is to design $\eta_k$ to make $[\rho_f, H_0'] = 0$. If the number of the control Haimltonians $r$ is large enough, by designing appropriate $\eta_k$, $[\rho_f, H_0'] = 0$ can be satisfied in most cases. Then one can design the control laws and analyze the convergence according to the method mentioned in Section 3.3. Research results show that every conclusion in Section 3.3 also holds with changing $H_0$ into $H_0'$.

## 4. Conclusion

In this paper, for the non-degenerate and degenerate cases, the existing quantum Lyapunov control based on the state distance, state error, and average value of an imaginary mechanical quantity for the control systems have been summarized and analyzed. For the target state being the eigenstate, the mixed state which commutes with the internal Hamiltonian, the superposition state, and the mixed state which does not commute with the internal Hamiltonian, respectively, the design methods of the control laws have been summarized; the convergence to the target state has been summarized and analyzed. Research results show that the Lyapunov-based quantum control method can make the control system converge from any initial state to the target state in both non-degenerate and degenerate cases. After ten years of development, the quantum control theory based on the Lyapunov stability theorem has been established.

## Acknowledgments

This work was supported in part by the National Key Basic Research Program under Grant no. 2011CBA00200 and the National Science Foundation of China under Grant no. 61074050.

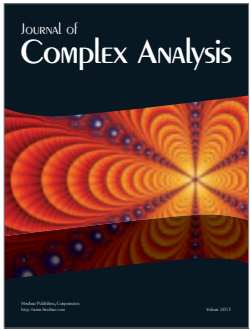

Journal of
**Complex Analysis**

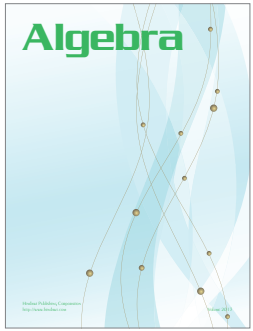

**Algebra**

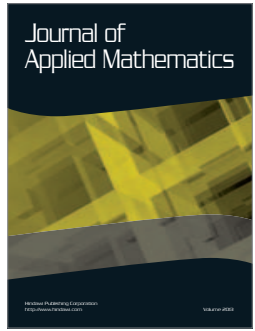

Journal of
Applied Mathematics

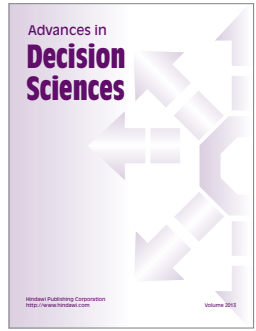

Advances in
**Decision Sciences**

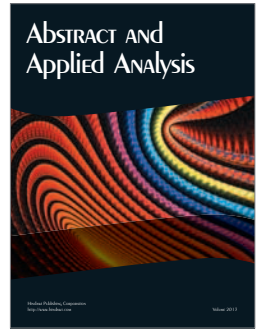

Abstract and
Applied Analysis

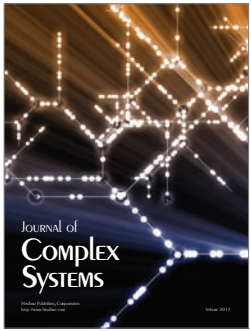

Journal of
**Complex Systems**

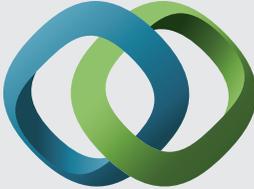

Hindawi

Submit your manuscripts at
http://www.hindawi.com

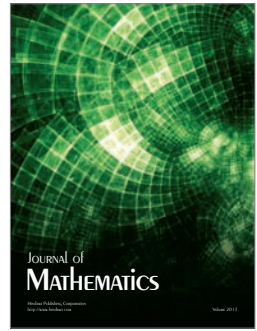

Journal of
**Mathematics**

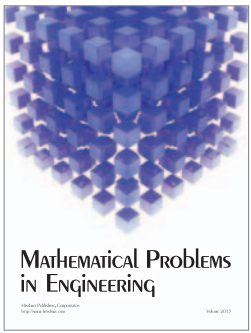

Mathematical Problems
in Engineering

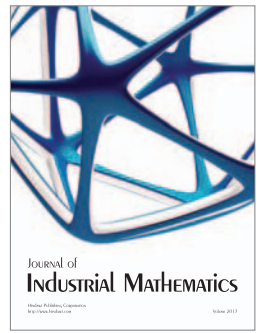

Journal of
**Industrial Mathematics**

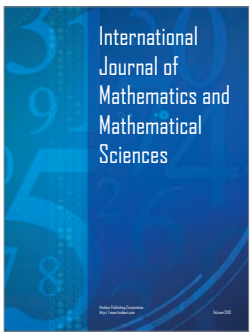

International
Journal of
Mathematics and
Mathematical
Sciences

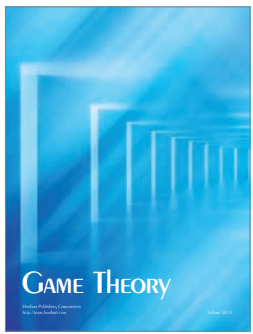

**Game Theory**

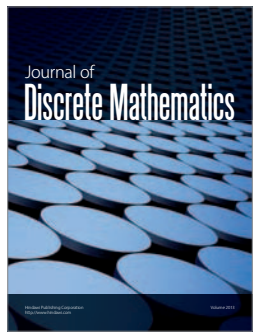

Journal of
**Discrete Mathematics**

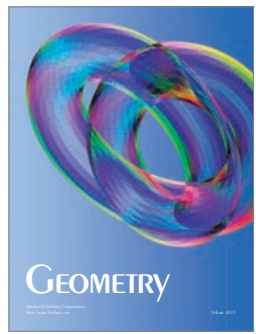

**Geometry**

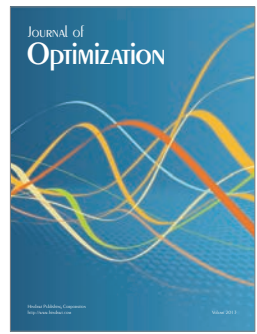

Journal of
**Optimization**

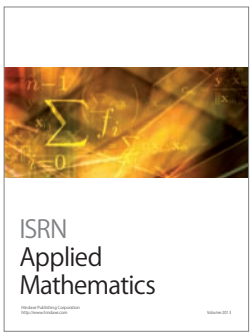

ISRN
Applied
Mathematics

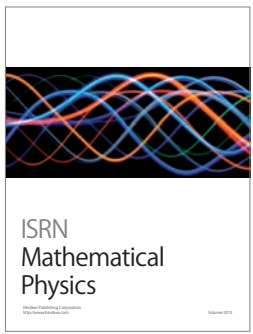

ISRN
Mathematical
Physics

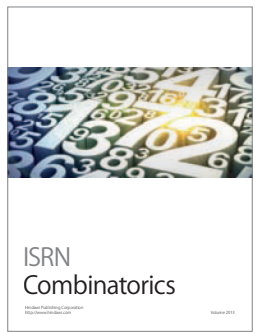

ISRN
Combinatorics

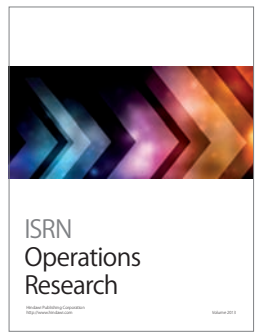

ISRN
Operations
Research

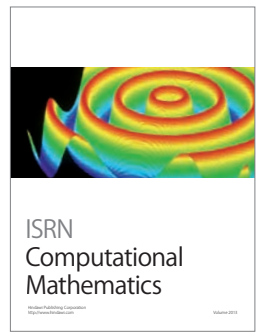

ISRN
Computational
Mathematics